\documentclass[5p,twocolumn]{elsarticle}
\usepackage{graphicx,latexsym}
\usepackage{dcolumn}
\usepackage{amssymb,amsmath,bm}
\usepackage{subfigure}
\usepackage{braket}
\usepackage{siunitx}
\usepackage{array}
\usepackage{float}
\usepackage[nopar]{lipsum}
\usepackage{caption}
\usepackage{multirow}
\usepackage{bigstrut}
\usepackage[super]{nth}

\newcommand{\angstrom}{\text{\normalfont\AA}}
\usepackage{hyperref}
\hypersetup{
    pdfnewwindow=true,       % links in new window
    colorlinks=true,         % false: boxed links; true: colored links
    linkcolor=blue,          % color of internal links
    citecolor=blue,          % color of links to bibliography
    filecolor=magenta,       % color of file links
    urlcolor=black           % color of external links
}

\usepackage[normalem]{ulem}

\def\sec#1{Sec.\ \ref{#1}}

\def\fig#1{Fig.\ \ref{#1}}
\def\tab#1{Tab.\ \ref{#1}}

\journal{}

\begin{document}

\begin{frontmatter}

%-----------------------------------------------------------------

\title{Enhanced electronic and optical responses of Nitrogen- or Boron-doped BeO monolayer: First principle computation}

\author[a1,a2]{Nzar Rauf Abdullah}
\ead{nzar.r.abdullah@gmail.com}
\address[a1]{Division of Computational Nanoscience, Physics Department, College of Science, 
             University of Sulaimani, Sulaimani 46001, Kurdistan Region, Iraq}
\address[a2]{Computer Engineering Department, College of Engineering, Komar University of Science and Technology, Sulaimani 46001, Kurdistan Region, Iraq}

\author[a3]{Botan Jawdat Abdullah}
\address[a3]{Physics Department, College of Science- Salahaddin University-Erbil, Erbil, Kurdistan Region, Iraq}

\author[a1]{Hunar Omar Rshid}

\author[a4]{Chi-Shung Tang}
\address[a4]{Department of Mechanical Engineering,
	National United University, 1, Lienda, Miaoli 36003, Taiwan} 

\author[a5]{Andrei Manolescu}
\address[a5]{Reykjavik University, School of Science and Engineering, Menntavegur 1, IS-101 Reykjavik, Iceland}

\author[a6]{Vidar Gudmundsson}
\address[a6]{Science Institute, University of Iceland, Dunhaga 3, IS-107 Reykjavik, Iceland}

%----------------------------------------------------------------

\begin{abstract}

In this work, the electronic and optical properties of a Nitrogen (N) or a Boron (B) doped BeO monolayer are investigated in the framework of density functional theory. It is known that the band gap of a BeO monolayer is large leading to poor material for optoelectronic devices in a wide range of energy. Using substitutional N or B dopant atoms, we find that the band gap can be tuned and the optical properties can be improved. In the N(B)-doped BeO monolayer, the Fermi energy slightly crosses the valence(conduction) band forming a degenerate semiconductor structure. The N or B atoms thus generate new states around the Fermi energy increasing the optical conductivity in the visible light region. Furthermore, the influences of dopant atoms on the electronic structure, the stability, the dispersion energy, the density of states, and optical properties such as the plasmon frequency, the excitation spectra, the dielectric functions, the static dielectric constant, and the electron energy loss function are discussed for different directions of polarizations for the incoming electric field. 

\end{abstract}

\begin{keyword}
BeO monolayer \sep DFT \sep Electronic structure \sep  Optical properties \sep Doping 
\end{keyword}

\end{frontmatter}

\section{Introduction}

Low dimensional nanomaterials have attracted the attention of scientists and technological experts in recent decades due to their extraordinary properties and superior performance compared to their bulk counterparts. Two-dimensional (2D) materials, in particular, are attractive owing to their atomic-thin layer structure and a variety of unique features. They still need extensive research and development to qualify for magnetic, electrical, optical, mechanical, and catalytic innovative applications 
\cite{Novoselov10451, doi:10.1021/acsnano.5b05556, ABDULLAH2020100740, WANG2021, Xiao_2019, GUPTA201544, RASHID2019102625}. 

Among the 2D materials, monolayer beryllium oxide (BeO), a new graphene-like metal oxide compound, has received a lot of attention.  
BeO is an alkali earth oxide with a wurtzite structure and a wide direct band gap, unlike most others, which have a cubic structure \cite{doi:10.1063/1.3075814}. Compton scattering experiments indicated that bonding in BeO is partially covalent with lower ionicity \cite{doi:10.1063/1.479262}. 
Several theoretical investigations on the structural determination of BeO have been undertaken through density functional theory (DFT). The unusual structural properties, such as the magnetic characteristics of B-, C-, and N-doped bulk BeO have been investigated, and the results reveal that the magnetic state of the sp-material is sensitive to charge and defect distribution \cite{Pang_2013}, the electrical and magnetic properties of BeO nanotubes \cite{PhysRevB.76.085407} and doped BeO nanotubes, are influenced by the type of impurities and their locations \cite{GORBUNOVA2008164}. The electrical and the magnetic characteristics of BeO nanoribbons may be altered by varying the edge morphologies and by passivation \cite{doi:10.1021/am201271j}.

\lipsum[0]
\begin{figure*}[htb]
	\centering
	\includegraphics[width=0.8\textwidth]{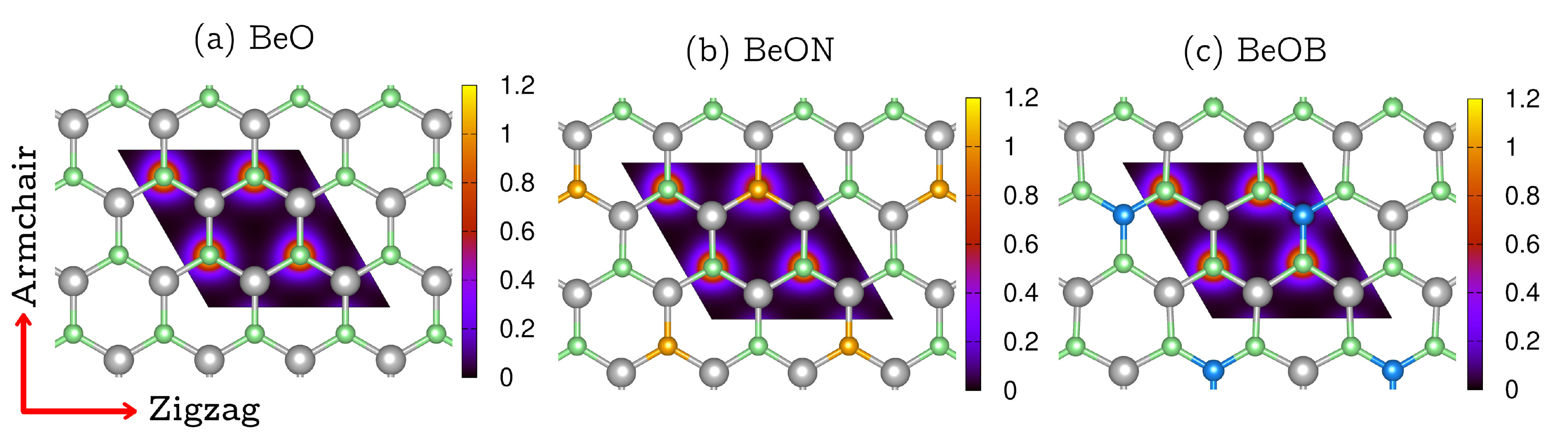}
	\caption{Position of atoms for optimized structures of BeO (a), BeON (b), and BeOB (c). 
		Be, O, N, and B atoms are gray, light green, orange, and blue color, respectively. The N atoms are doped at Be atom position forming BeON-1 monolayer while the B atoms are put at the O atom position forming BeOB-2 monolayer.}
	\label{fig01}
\end{figure*}

Using in-situ analytical methods, a thin BeO layer has been produced experimentally \cite{Oberkofler2011}, 
and hexagonal BeO monolayer has recently been synthesized via molecular epitaxial growth on an Ag(1 1 1) thin film \cite{doi:10.1021/acsnano.0c06596}.
Theoretically, a monolayer BeO exhibits great dynamic stability \cite{PhysRevB.92.115307} as well as thermal, kinetic, and mechanical stability due to atomic hybridization and electronic delocalization \cite{Luo17213}. 

DFT and the Boltzmann transport equation have been used to investigate the electrical and thermal characteristics of a BeO monolayer and found that, due to the ionic nature of the Be-O bond, electrons in a monolayer BeO are strongly concentrated around the O and the Be atoms, which differs from the situation in graphene and hexagonal boron nitride. Additionally, the thermal conductivity of a 2D monolayer BeO is 266 W/mK at 300 K and drops with increasing temperature, which is mostly due to low frequency phonons \cite{Xia_2020}.

A limited number of theoretical investigations of BeO monolayers have previously reported that this material is very promising for novel properties. A remaining challenge in the doping. For example, density functional theory simulations were used to investigate the electrical and magnetic characteristics of monolayer BeO with transition metal (TM) substitutional doping. The findings indicate that the electrical and the magnetic characteristics of monolayer BeO with different TM substitutional can be tuned \cite{SONG2018252}, and the magnetic behavior of an N-doped BeO monolayer was induced by the spin-up and spin-down band gaps, which depend on the dopant concentration and the N–N separation \cite{Hoat_2021}.

As a result, the impact of impurity atoms on the fundamental characteristics of BeO monolayers have yet to be described, as they may be useful for novel applications in nanotechnology. As a consequence, monolayer BeO doped with N and B forming BeO$_x$N$_{1-x}$ and Be$_x$OB$_{1-x}$, respectively, is studied in this work. Using DFT models, we show how doping may be used to tune the electrical and optical properties of a BeO monolayer. BeO monolayer doped with N and B atoms have distinct band structures depending on the dopant configuration. According to the results for the excitation spectra, the dielectric function, the electron energy loss function, and the optical conductivity, we will see that the N- or B-doped BeO monolayer material is suitable for optoelectronic applications.

In \sec{Sec:Model} the computational techniques and the model structure are briefly overviewed. In \sec{Sec:Results} the main achieved results are analyzed. In \sec{Sec:Conclusion} the conclusion of results is presented.

\section{Computational details}\label{Sec:Model}

Electronic and optical properties of pure BeO, B- or N-doped BeO monolayers are studied within first principle full potential projector augmented wave method (PAW) via the DFT
as implemented in Quantum Espresso, QE, code \cite{Giannozzi_2009, giannozzi2017advanced}. 
We have made use of the GGA for the exchange-correlation functional, the projector augmented
wave method, and the plane-wave basis set with an energy cutoff $1088$~eV \cite{ABDULLAH2021106073, ABDULLAH2020126807}. 

The Brillouin zone sampling is done for the supercell with the equivalent of a $15\times15\times1$
Monkhorst-Pack $k$-point grid for a BeO unit cell (containing one Be and one O atom). 
The atomic positions are considered relaxed when the forces are smaller than $0.001$ eV/$\angstrom$. 
To decrease the strain caused by the substituents, the lattice parameters
are very carefully optimized \cite{ABDULLAH2021413273}.
In addition, the partial occupancies in the electronic ground-state, and the density of state, DOS, calculations are treated by using Gaussian smearing, and the Tetrahedron methodology \cite{PhysRevB.49.16223, ABDULLAH2021114644, PhysRevB.13.5188}, respectively. In the DOS calculations, a dense $100\times100\times1$ grid is used in order to obtain very smooth DOS curves with accurate peak positions.
To suppress the interaction emerging from periodic boundary conditions in the $z$-direction, a $2\times2\times1$ supercell is used with the height of $20 \, \angstrom$ to include enough vacuum  around the layer \cite{ABDULLAH2021106981}.
The optical properties of the BeO systems are calculated by the QE code, and 
an optical broadening of $0.1$~eV is assumed for the calculation of
the dielectric properties \cite{ABDULLAH2021110095, ABDULLAH2020126578}.

Finally, the crystalline and molecular structure visualization software (XCrySDen) and VESTA
are utilized in this work to visualize the structures \cite{KOKALJ1999176, momma2011vesta}.

\section{Results}\label{Sec:Results}

We first examine the structural properties of pure BeO, and B- or N-doped BeO monolayers. 
The BeO monolayer is one atom thick, with a flat hexagonal honeycomb lattice
like graphene, silicene, and BN monolayers, exclusively made of Be-O bonds.
Since a unit cell of BeO consists of one Be and one O atoms, the B- or the N-doped BeO can have two substitutional doping possibilities. These two doping configurations are expected to give different physical properties because of the different electronegativity and atomic radii of the Be and O atoms. First: an N or B atom is substitutionally doped at a Be atom position of the BeO monolayer identified as BeON-1 or BeOB-1 monolayer, respectively. 
Second, an N or B atom is substitutionally doped at an O atom position of a BeO monolayer identified as BeON-2 or a BeOB-2 monolayer, respectively. 
In \fig{fig01} the monolayers of pure BeO (a), BeON (BeON-1) (b) and BeOB (BeOB-2) (c) are presented 
with their electron charge distribution shown in the parallelogram.

We assume a $2\times2\times1$ supercell monolayer where the dopant ratio is assumed to be $12.5\%$. 
Since the electronegativity of the O atom, $3.44$, is much higher than that of the Be atom, $1.57$, the electron charge distribution around the O atoms in the hexagon of the monolayer is much higher than that of the Be atoms, as it is seen in \fig{fig01}(a). A strong electron charge distribution around the N atom is seen while a weak electron charge distribution around the B atom is recorded. The strong charge localization of the N atom is referred to its electronegativity which is close to the electronegativity of the O atom.

The PBE optimized lattice constants and Be-O bond length are found to be 
$2.67 \, \angstrom$, and  $1.54 \, \angstrom$, respectively, for pure BeO monolayer. These results agree very well with recent theoretical \cite{MORTAZAVI2021100257} and experimental \cite{doi:10.1021/acsnano.0c06596} studies of BeO monolayers. Information about the lattice constants and bond lengths of N- or B-doped BeO monolayers are shown in 
\tab{table_one}. In general, the average value of the lattice constant of an N- or B-doped BeO monolayer is increased indicating a supercell expansion due to the interaction between the N- or B- atoms with the Be and O atoms.
Consequently, the average Be-O bond length of BeON and BeOB is also elongated compared to a pure BeO monolayer.

\begin{table}[h]
	\centering
	\begin{center}
		\caption{\label{table_one} Lattice constant, $\it a$, Be-O, Be-N, Be-B, O-N, O-B bond lengths for all structures under investigation. The unit of all parameters is $\angstrom$.}
		\begin{tabular}{|l|l|l|l|l|l|l|}\hline
			Structure	  & $ \it a$     & Be-O    & Be-N & Be-B  & O-N   & O-B     \\ \hline
			BeO	          & 2.67  & 1.54    &  -   & -     & -     & -       \\
			BeON-1	      & 2.885 & 1.598   &  -   & -     &1.408  & -       \\			BeON-2	      & 2.731 & 1.556   &1.608 & -     &  -    & -       \\
			BeOB-1	      & 2.6   & 1.558   &  -   & -     &  -    & 1.416   \\ 
			BeOB-2	      & 2.89  & 1.584   &  -   & 1.824 &  -    & -       \\ \hline
		\end{tabular}  
	\end{center}
\end{table}

We calculate the formation energy of all the structures. Formation energy is the energy required for generating the atomic configuration of the structure, and can be used as an indication for the
energetic stability of the N- or B-doped BeO monolayers \cite{ABDULLAH2020114556}. Our PDF calculations indicate that the formation energy of the four types of doped BeO monolayers can be arranged from lower to higher formation energy as follow: BeON-1 $<$ BeOB-1 $<$ BeON-2 $<$ BeOB-2. The lower the formation energy, the more energically stable structure is obtained. As a result, the BeON-1 and BeOB-1 are the most stable structures. We note that the structure is more stable if a dopant atom is put at the Be position of the hexagon of the BeO monolayer.

\subsection{Nitrogen doped BeO monolayer}

We now study the electronic band structure and the DOS, and optical properties of BeON monolayer, with the two types of the BeON monolayers considered.
\begin{figure}[htb]
	\centering
	\includegraphics[width=0.48\textwidth]{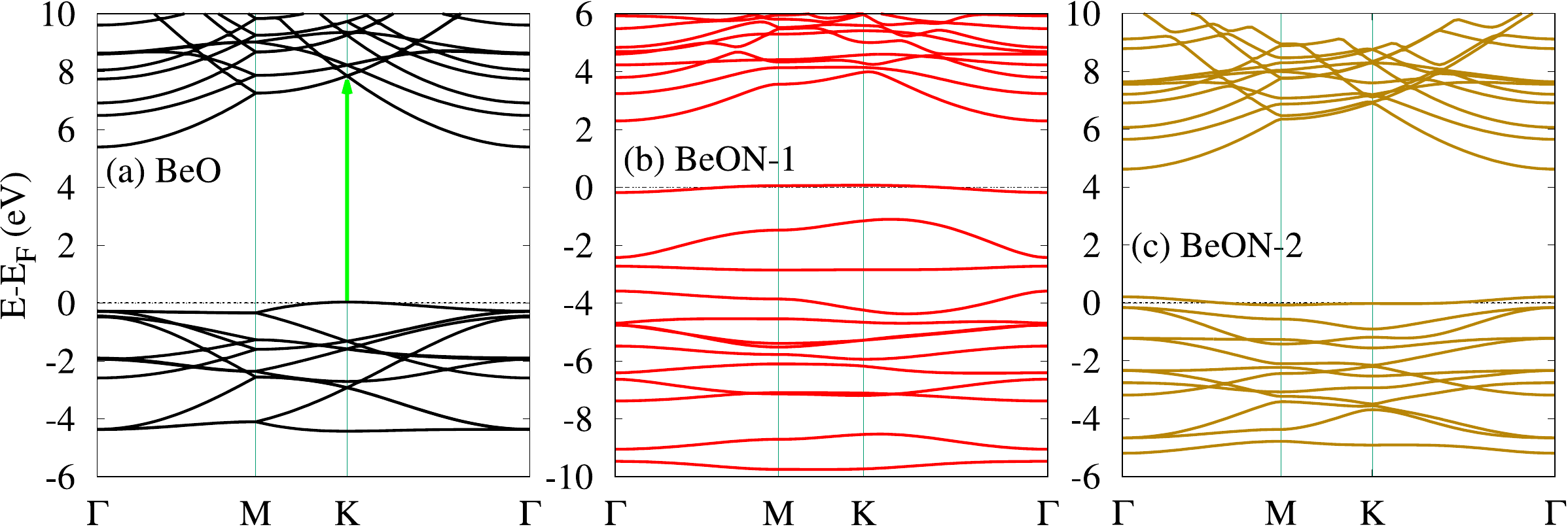}
	\caption{Band structure for optimized structures of BeO (a), BeON-1 (b), and BeON-2 (c). 
		The energies are with respect to the Fermi level, and the Fermi energy is set to zero.}
	\label{fig02}
\end{figure}
The band structure of BeO (a), BeON-1 (b), and BeON-2 (c) are shown in \fig{fig02}. The results for a pure BeO monolayer are shown here to compare with the doped BeO monolayers.
An experimental observation indicates that BeO is an insulator with a band gap of $6.4$~eV.
In our calculations, using PBE functionals to calculate the electronic band structures of the BeO monolayers along the high symmetry directions of its Brillouin zone.
We find a band gap of BeO monolayer to be $5.4$~eV, which is very close to the theoretical results using PBE functionals in a recent work \cite{MORTAZAVI2021100257}, while in calculation using the HSE06 functionals the band gap is $6.7$~eV, which indicates that the band gap shown in \fig{fig02} are underestimated. It is also found that the band gap of a BeO monolayer is an indirect band gap displaced from the K- to the $\Gamma$-point. 

\begin{figure}[htb]
	\centering
	\includegraphics[width=0.45\textwidth]{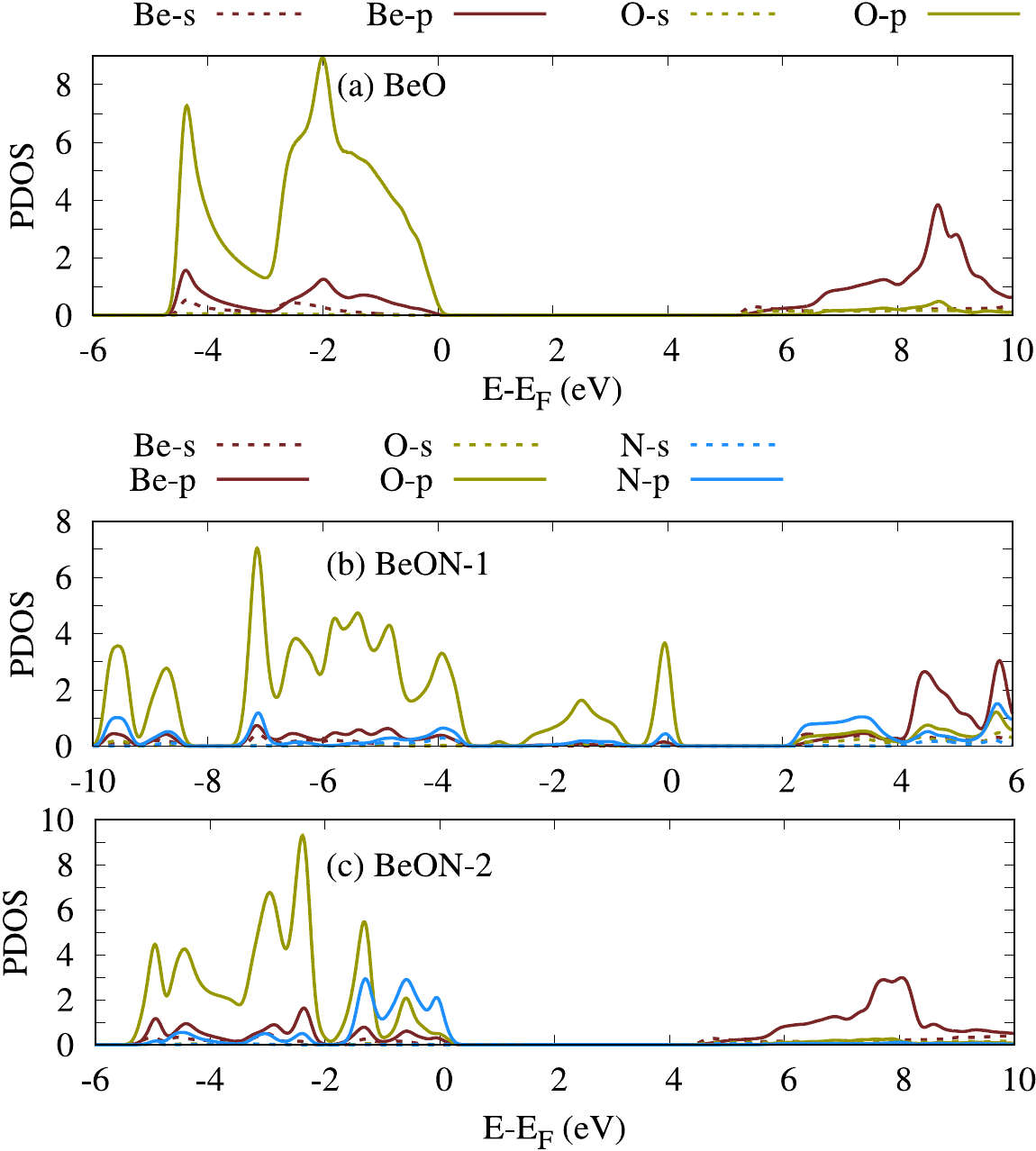}
	\caption{Partial density of states (PDOS) of pure BeO (a), BeON-1 (b), and BeON-2 (c). The PDOS of s- (dashed lines) and the p-orbital (solid lines) of all three atoms (Be, O, and N) are plotted. The Fermi energy is set to zero.}
	\label{fig03}
\end{figure}

\lipsum[0]
\begin{figure*}[htb]
	\centering
	\includegraphics[width=0.8\textwidth]{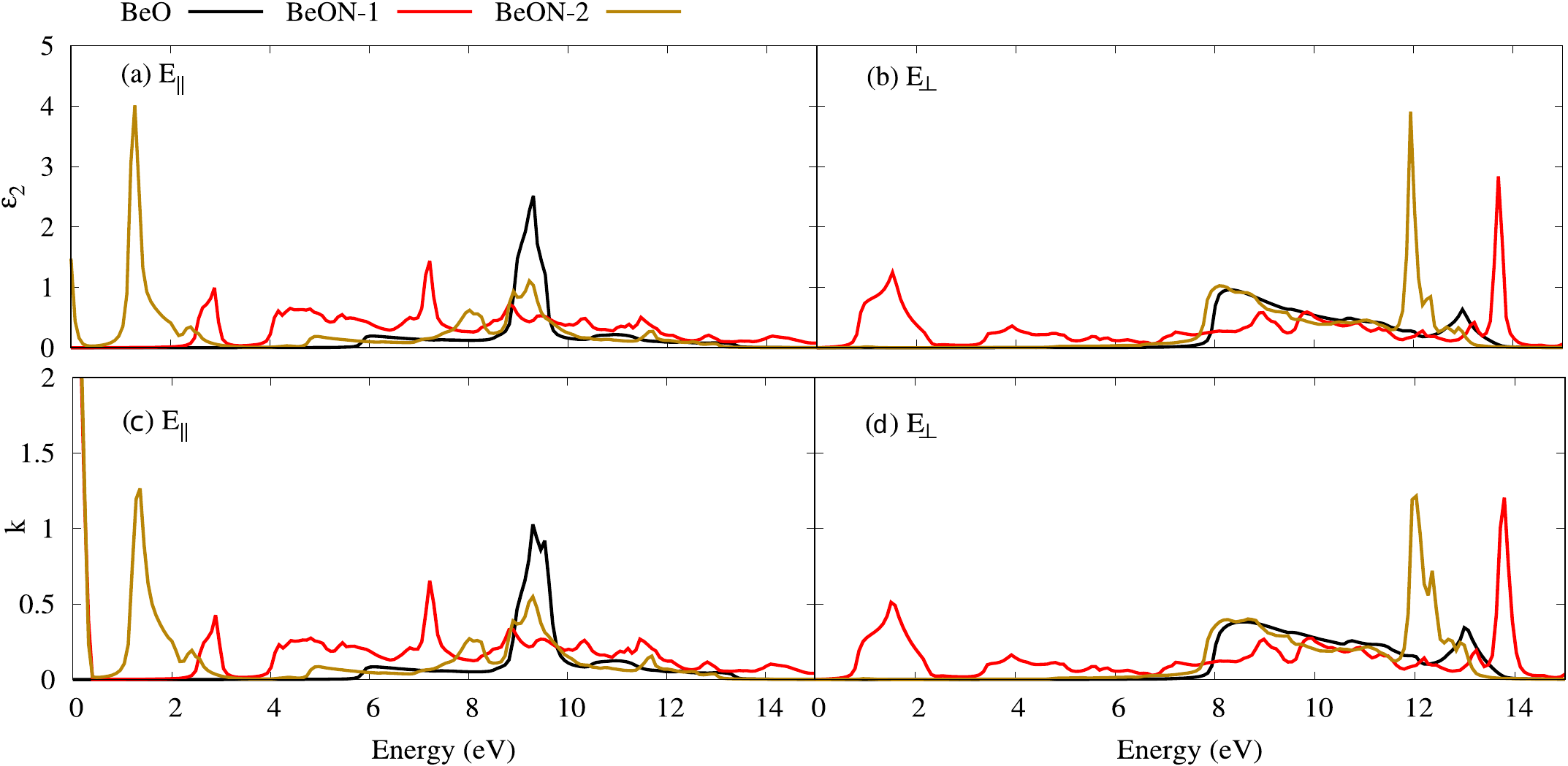}
	\caption{Imaginary part of the dielectric function, $\varepsilon_2$ (a) and (b), and excitation spectra, $k$ (c) and (d), for a parallel $E_{\rm \parallel}$ (left panel) and perpendicular $E_{\rm \perp}$ (right panel) polarizations of the electric field, respectively.}
	\label{fig04}
\end{figure*}

To get insight into the band structure, we plot the partial density of states, the PDOS, of pure BeO (a), BeON-1 (b), and BeON-2 (c) monolayers in \fig{fig03}.

We can see that the valence bands of BeO are formed by hybridization of $p$-orbitals of both the Be and O atoms, where the O atoms account for the main contributions. 
In contrast, in the formation of the conduction bands of BeO, the $p$-orbital of the
Be atoms play a major role. The valence band maxima and the conduction band minima are mainly due to the $p_z$-orbitals of both the O and Be atoms, respectively (not shown).

In an N-doped BeO monolayer, the conduction band shape is almost unchanged, while 
the valence bands are strongly modified. The modification of the valence bands of BeON-1 is stronger than that of BeON-2. This is caused by the N atom doped in the position of a Be atom position in the BeON-1, as the atomic radii of an N atom is much smaller than that of a Be atom.
It thus leads to a stronger symmetry breaking. But the modification in BeON-2 is weaker as the N atom is doped at a postition of an O, and the atomic radii of both an N and an O atom are very close to each other. Consequently, only slight modification in the valence band is seen for a BeON-2 monolayer.

Most importantly, similar to a N-doped graphene monolayer \cite{ABDULLAH2020126350}, the N dopant atom slightly shifts the Fermi energy to the valence bands in both BeON-1 and BeON-2 monolayers. 
The Fermi level only slightly enters the valence bands accounting for degenerate semiconductor properties of both BeON-1, and BeON-2 monolayers.
The contribution of the N atom in the band structure can further be confirmed by showing the PDOS of 
BeON-1 (b) and BeON-2 (c) in \fig{fig03}. 
In BeON-1, the influence of the N dopant atom is more effective in the conduction band region leading to generation of new states between $2.0\text{-}4.0$~eV, which is expected because an N atom is doped at the position of Be atoms. As a result, the conduction bands shift down in addition to the shifting up of the valence band which decrease the band gap of BeON-1 as shown in \fig{fig02}(b). We note that the valence bands crosses the Fermi energy along the M-K direction in BeON-1.

In contrast to the BeON-1, the N dopant atoms contribute to the valence band in the BeON-2, where the N atom is doped at the O atom position. The new states due to the N atom appear in the energy range from $0.2$ to $-2.0$~eV as it is seen in the PDOS of BeON-2, \fig{fig03}(c)(blue). As a result, the valence band crossing the Fermi energy is seen in the energy spectrum shown in \fig{fig02}(c), and a reduction in the band gap is also found. The valence bands cross the Fermi energy near the $\Gamma$ points in BeON-2. The effects of the N dopant atom in the conduction band region is very small, and can be ignored as the difference between the conduction bands of BeON-2 monolayer and a pure BeO monolayer is vanishingly small.  

Our next aim is to study the optical properties of both BeON-1 and BeON-2 monolayers, in addition to pure BeO monolayer. The real, $\varepsilon_1$, and the imaginary, $\varepsilon_2$, parts of the dielectric function can be obtained from the QE software using an RPA approximation \cite{ABDULLAH2021106073}.
The imaginary part of the dielectric function, and the excitation spectra are presented in \fig{fig04} for parallel, $E_{\rm \parallel}$, (a, c) and perpendicular, $E_{\rm \perp}$, (b, d) polarization of electric fields with respect to the surface of monolayers.

In the pure BeO monolayer, the threshold energies for $E_{\rm \parallel}$ and $E_{\rm \perp}$ derived from $\varepsilon_2$ are $5.85$ and $7.81$~eV, respectively. In addition, a peak in both  $\varepsilon_2$, and $k$ are found at the $9.3$~eV for $E_{\rm \parallel}$ indicating a transition from the valence band maxima to the conduction band at the K point as is shown in \fig{fig02}(a) (light green arrow).
We find that the properties of $\varepsilon_2$ for the BeO monolayer are completely different from the dielectric function for BeO bulk in wurtzite structure \cite{Valedbagi2014, Amrani_2007}.
The difference between the monolayer and the wurtzite structure is due to differences in the Be-O hybridization. 
In a BeO monolayer each atom (Be or O atoms) has three bonds with nearest neighbor, but 
in the BeO wurtzite structure each atom is bonded with four other atoms.

In the $\varepsilon_2$ and $k$ characteristics for BeON-1 (red) and BeON-2 (golden), several peaks revealing transitions from the valence to the conduction bands are found for both polarizations of electric field. 
More importantly, peaks in the low energy range between $0\text{-}2$~eV in the visible regime are seen. These peaks are caused by the degenerate semiconductor characteristics of both BeON-1 and BeON-2, caused by the slight crossing of the Fermi level into the valence band. We note that the peaks at low energy are more pronounced for BeON-2 for $E_{\rm \parallel}$, while 
they are stronger for BeON-1 for $E_{\rm \perp}$. This again can be referred to the different location of the N dopant atoms in these monolayers.

\begin{figure}[htb]
	\centering
	\includegraphics[width=0.45\textwidth]{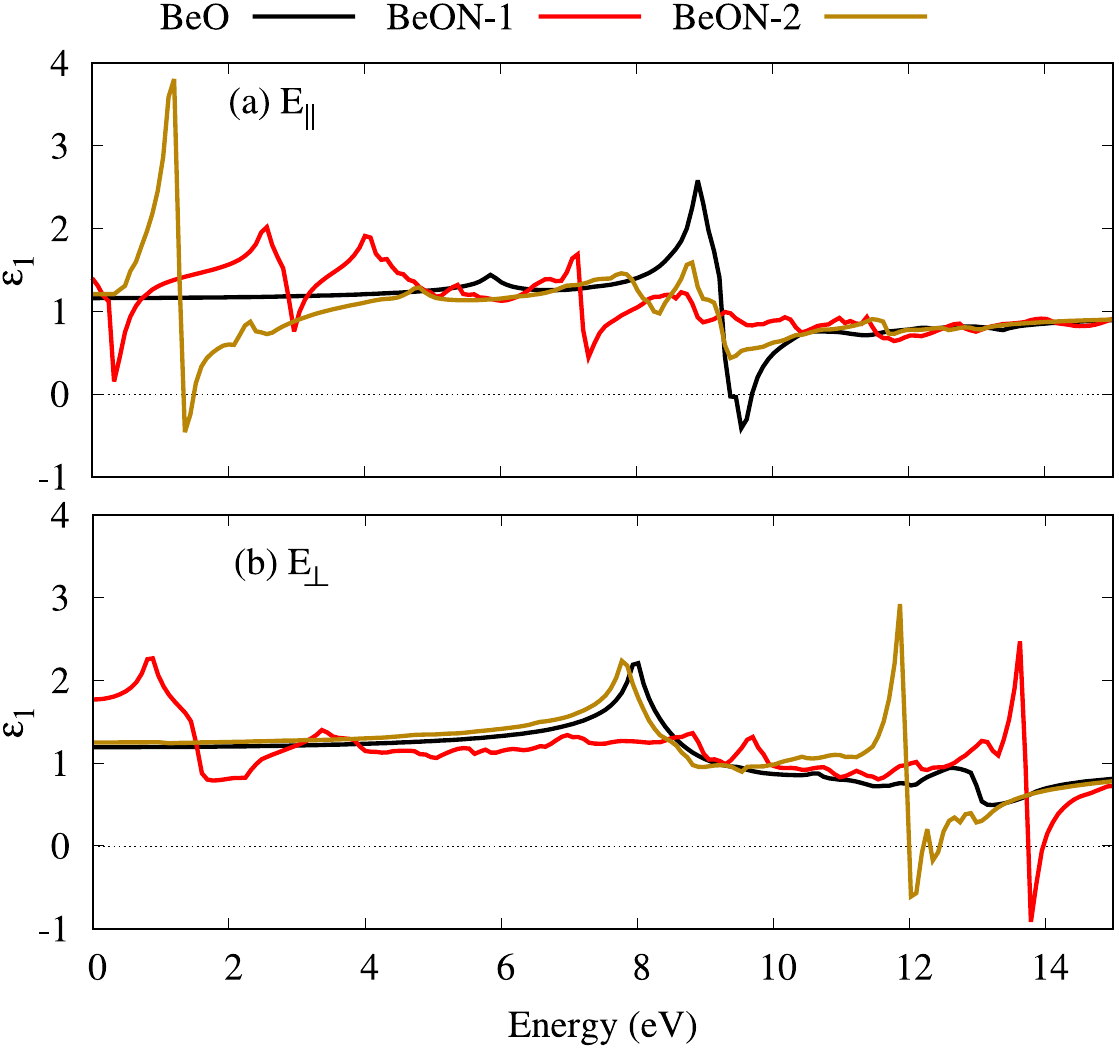}
	\caption{Real part of the dielectric function, $\varepsilon_1$, for BeO (black), BeON-1 (red), and BeON-2 (orange) in $E_{\rm \parallel}$ (a) and $E_{\rm \perp}$ (b).}
	\label{fig05}
\end{figure}

The real part of the dielectric function, $\varepsilon_1$, is also an important parameter for a material, as it serves like an indication of the degree to which the material can be polarized. The real parts of the dielectric function of BeO,  BeON-1, and BeON-2 monolayers are demonstrated in \fig{fig05} for $E_{\rm \parallel}$ (a) and $E_{\rm \perp}$ (b). We see that the values of the static dielectric constant for pure BeO, the value of the  dielectric constant at zero energy ($\varepsilon_1(0)$), is $1.16$ and $1.19$ for $E_{\rm \parallel}$ and $E_{\rm \perp}$, respectively.
This value is modified and enhanced for BeON-1 and BeON-2. In the presence 
of $E_{\rm \parallel}$, the value of $\varepsilon_1(0)$ becomes to $1.39$ and $1.207$ for BeON-1 and BeON-2, respectively. In addition, in the presence of $E_{\rm \perp}$, these values are changed to 
$1.77$ and $1.24$ for BeON-1 and BeON-2, respectively.

Another important physical parameter of the monolayers that can be extracted from the $\varepsilon_1$ spectra is the plasmon frequency, where the plasmons are the collective oscillations of the free electron density. At the plasmon frequency, the value of $\varepsilon_1$ should be zero (the frequency where $\varepsilon_1$ changes its sign from positive to negative or vise versa), and the value of 
$\varepsilon_2$ attains a maximum value. Based on the aforementioned conditions, the plasmon frequency for pure BeO monolayer is found to correspond to $9.3$~eV for $E_{\rm \parallel}$, while no plasmon mode is seen for the $E_{\rm \perp}$. 
It is interesting to see that the zero value of $\varepsilon_1$($\omega$) corresponding to the location of the screened plasma frequency appears for BeON-2 at $1.29$~eV for $E_{\rm \parallel}$, but for BeON-1 and BeON-2 in $E_{\rm \perp}$ they are observed at $13.75$ and $12.0$~eV, respectively.

\begin{figure}[htb]
	\centering
	\includegraphics[width=0.42\textwidth]{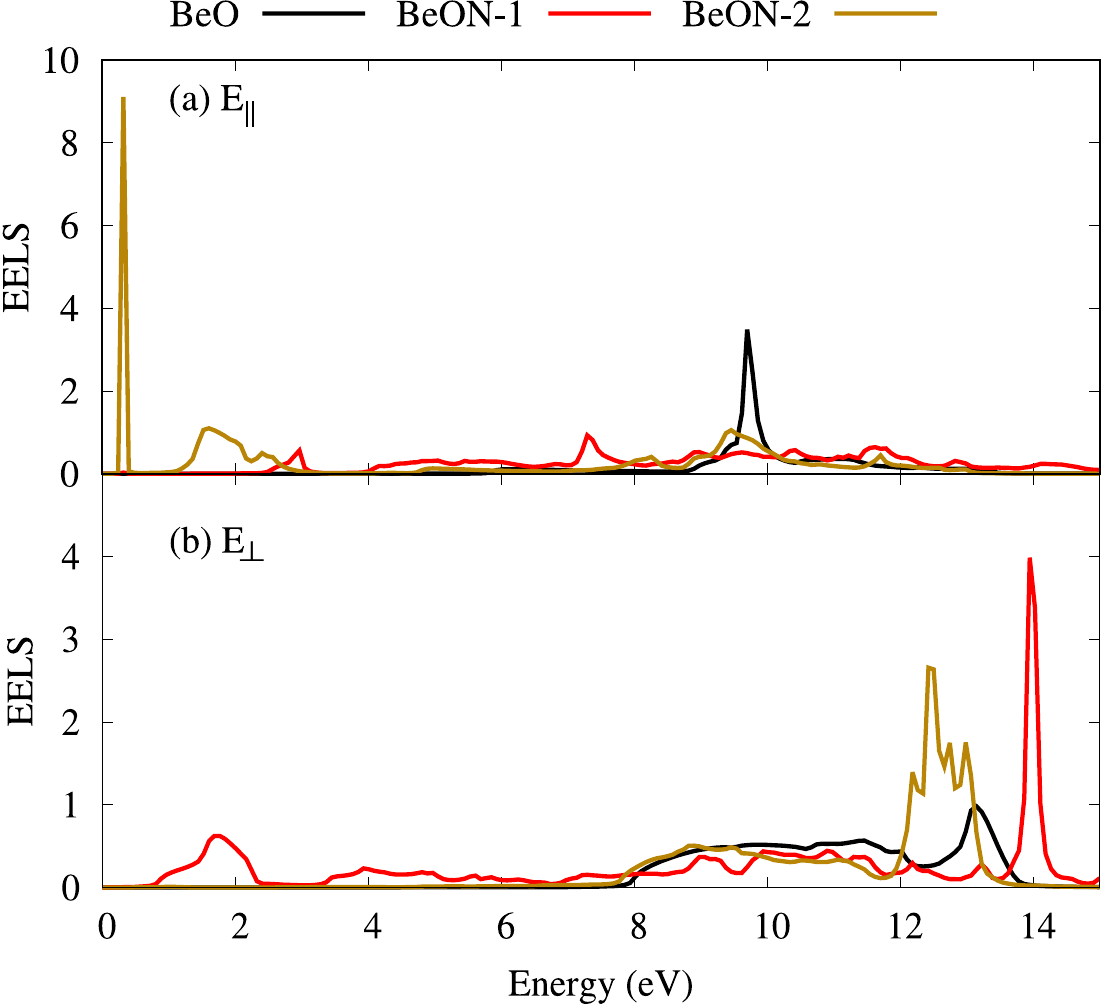}
	\caption{Electron energy loss spectrum, EELS, for BeO (black), BeON-1 (red), and BeON-2 (orange) in the case of $E_{\rm \parallel}$ (a) and $E_{\rm \perp}$ (b) electric field.}
	\label{fig06}
\end{figure}

The next targets are the electron energy loss spectra, EELS, of pure BeO, BeON-1, and BeON-2 monolayers for both $E_{\rm \parallel}$ (a) and $E_{\rm \perp}$ (b), which are shown in \fig{fig06}. 
The EELS demonstrates the energy lost in interaction process of fast moving electron traveling through the material. The interaction may arise via multiple processes of interband and intraband transitions, 
plasmon excitations, phonon excitation, and inner shell ionizations in the materials. 

In the selected range of energy, a sharp peak in the EELS of pure BeO monolayer at $9.7$~eV for $E_{\rm \parallel}$ and a broad peak at $13.15$~eV for $E_{\rm \perp}$ are seen. The peak at $9.7$~eV is clearly related to the $\pi$ electrons plasmon as the plasma frequency of BeO is found close to that energy value. Our results of EELS for BeO monolayer agree very well with a previous study \cite{FATHALIAN20131}.

The intense EELS peak for BeON-1 monolayer at $13.95$ in the $E_{\rm \perp}$, and the two peaks of BeON-2 monolayer at $1.6$ and $12.48$ for $E_{\rm \parallel}$ and $E_{\rm \perp}$, respectively, are caused by plasmon oscillations.   
In addition, clear peaks due to interband and intraband transitions at low energy are found for a BeON-2 monolayer for $E_{\rm \parallel}$.
Last, but not least about the EELS of BeON monolayers, one can see intense EELS peaks for BeON-2 in the $E_{\rm \perp}$ occurs at lower energy than for the BeON-1. This is related to the atomic configuration of both structures. In the BeON-1, the N atoms are doped at the positions of Be atoms, while in the BeON-2 monolayer the N atoms are put at O atom positions. The valence/conduction electron charge distribution in the BeON-1 is thus increased while it is decreased in the BeON-2. 
The higher the electron charge distribution of a structure, the more energy is needed to excite the collective oscillation or the plasmons. As a result, the intense plasmon peak in BeON-1 monolayer occurs at higher energy.

\begin{figure}[htb]
	\centering
	\includegraphics[width=0.45\textwidth]{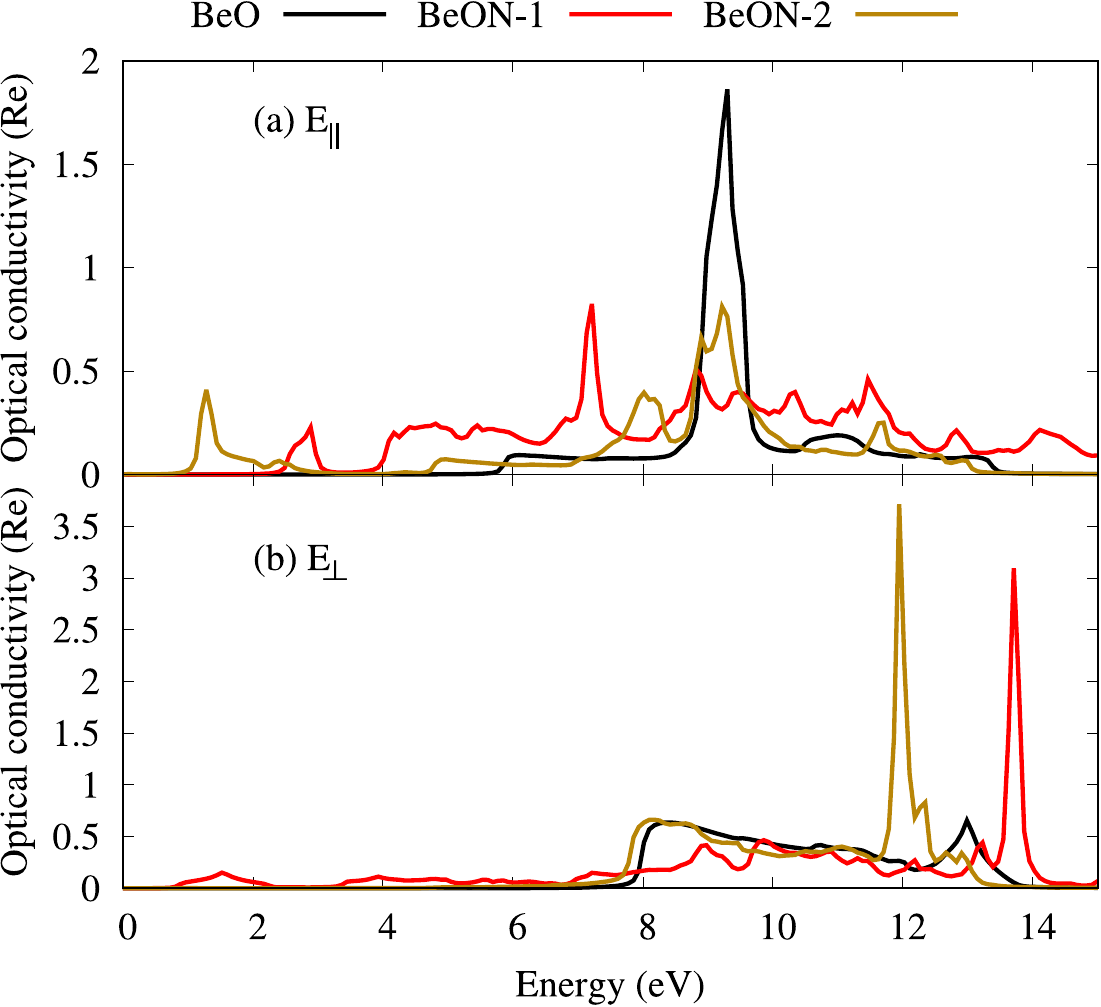}
	\caption{Optical conductivity (Real part), for BeO (black), BeON-1 (red), and BeON-2 (orange) in the case of $E_{\rm \parallel}$ (a) and $E_{\rm \perp}$ (b) electric field.} 
	\label{fig07}
\end{figure}

The real part of the optical conductivity is another interesting optical property of a material. 
The real part of the optical conductivity is related to the imaginary part of the dielectric function, 
$Re[\sigma(\omega)] = \omega \, \varepsilon_2(\omega)/4 \pi$. The real part of the optical conductivity
for BeO, BeON-1, and BeON-2 monolayers are presented in \fig{fig07} for both $E_{\rm \parallel}$ (a) and $E_{\rm \perp}$ (b) electric fields. It is clearly seen that the $Re[\sigma(\omega)]$ follows almost the same behavior as $\varepsilon_2$ for both directions of electric field polarization.

The optical conductivity of BeO in $E_{\rm \parallel}$ and $E_{\rm \perp}$ begins with a gap about $5.75$ eV and $7.6$ eV, confirming that a BeO monolayer hase semiconductor and insulator properties for  $E_{\rm \parallel}$ and $E_{\rm \perp}$, respectively. In both the BeON-1, and BeON-2 monolayers, several small peaks appear indicating inter- and intraband transitions at low energy range.
%-----------------------------------------------------------------

\subsection{Boron doped BeO monolayer}

\begin{figure}[htb]
	\centering
	\includegraphics[width=0.45\textwidth]{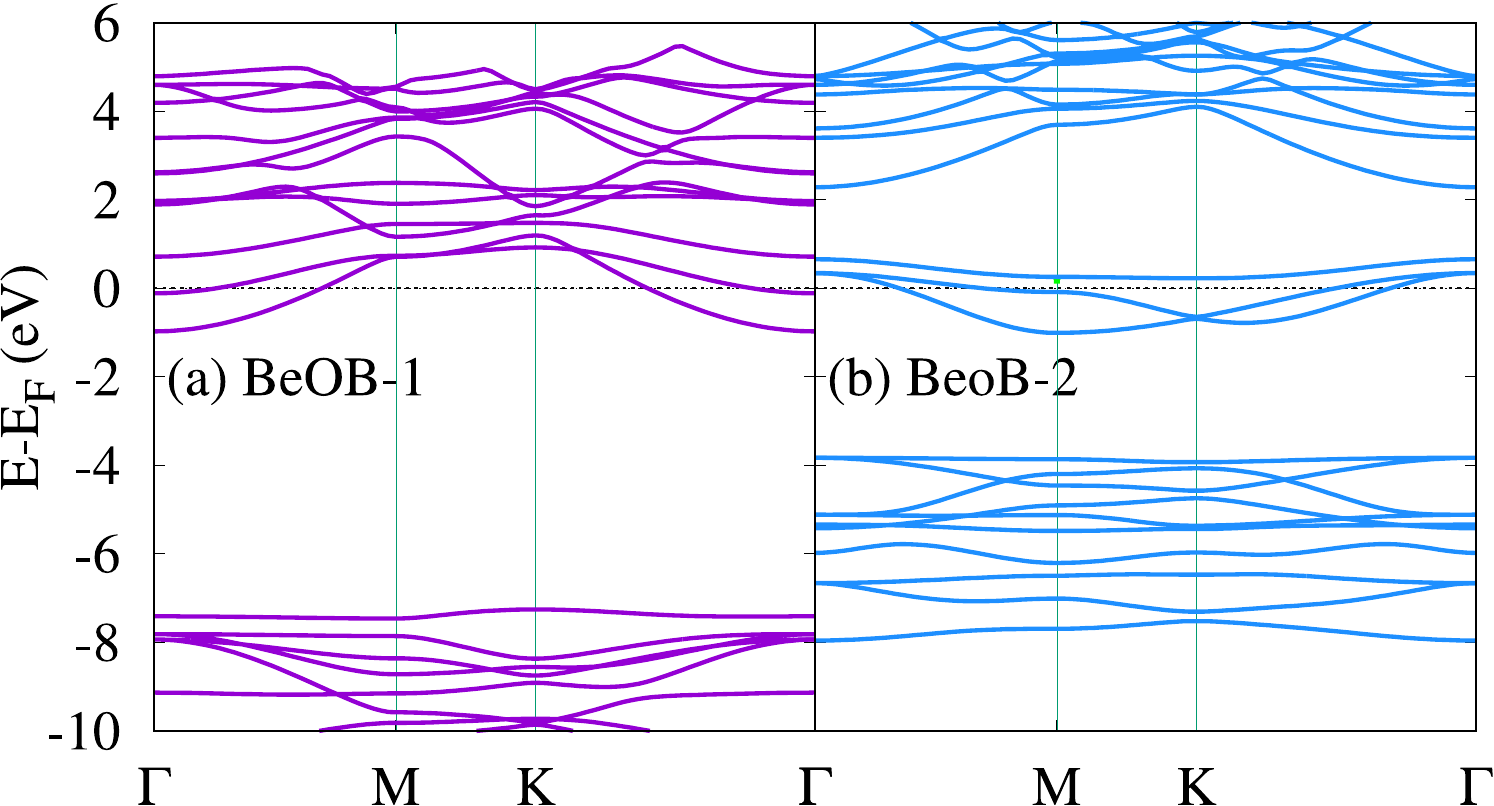}
	\caption{Band structure for optimized structures of BeOB-1 (a), and BeOB-2 (b). 
		The energies are with respect to the Fermi level, and the Fermi energy is set to zero.}
	\label{fig08}
\end{figure}

In this section, we consider a B doped BeO monolayer forming BeOB-1 with a B atom doped 
at the position of a Be atom, and BeOB-2 with a B atom put at the position of an O atom. 
The band structures of BeOB-1 (a) and BeOB-2 (b) are presented in \fig{fig08}.
Similar to B-doped graphene \cite{ABDULLAH2020126350}, the B atom shifts the Fermi energy into the conduction band, an opposite behavior to the N-doped BeO monolayer studied in the previous section.
The Fermi energy slightly crosses the conduction band region, and the crossing is near the $\Gamma$ point in BeOB-1, but along the M-K path in BeOB-2 monolayer. Again, the B-doped BeO monolayer acquiers properties of a degenerate semiconductor.

\begin{figure}[htb]
	\centering
	\includegraphics[width=0.48\textwidth]{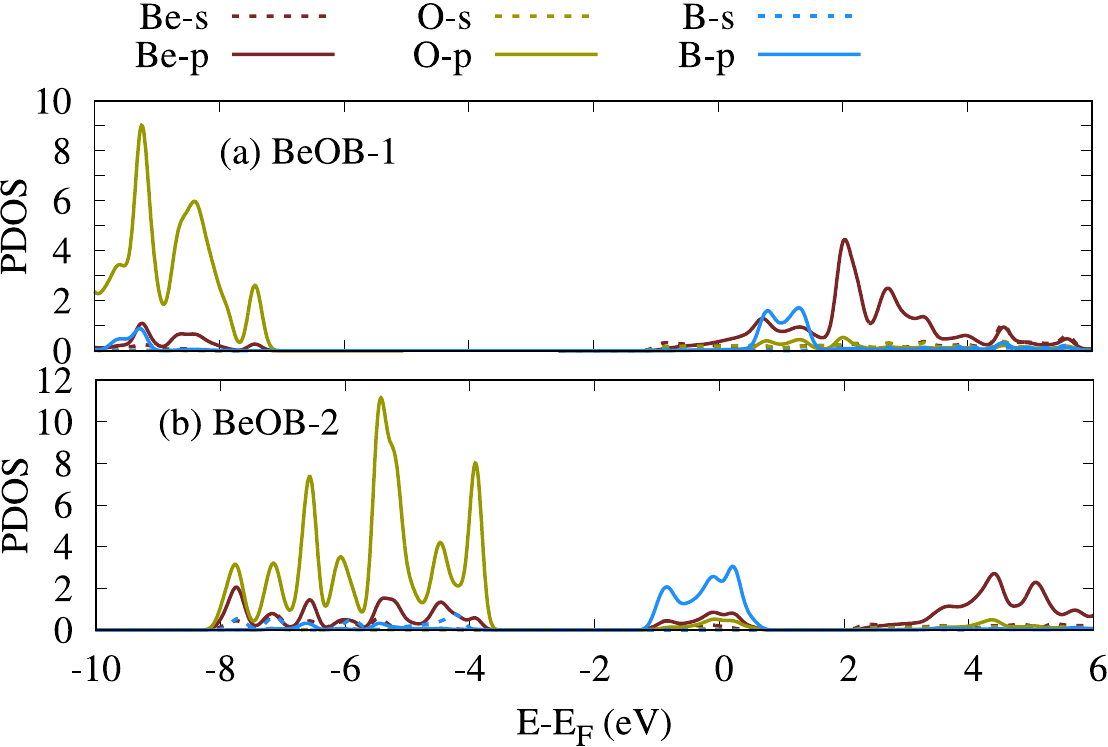}
	\caption{Partial density of states (PDOS) of the BeOB-1 (a), and BeOB-2 (b). The PDOS of s- (dashed lines) and the p-orbital (solid lines) of all three atoms (Be, O, and B) are plotted. The Fermi energy is set to zero.}
	\label{fig09}
\end{figure}

In order to better understand the nature of the band structure of a B-doped BeO, we plot the PDOS of both BeOB-1 (a) and BeOB-2 (b) in \fig{fig09}. It is obvious that the B atoms generate new states in the conduction band region near the Fermi energy. The contribution of $p$-orbital ($p_z$-orbital) of the B atom is dominant from $0.6$ to $1.6$~eV for BeOB-1, and from $-1.26$ to $0.8$~eV for BeOB-2.

The optical properties of a B-doped BeO monolayer such as the imaginary and the real parts of the dielectric function, the excitation spectra, EELS, and the optical conductivities are of interest. 
\lipsum[0]
\begin{figure*}[htb]
	\centering
	\includegraphics[width=0.8\textwidth]{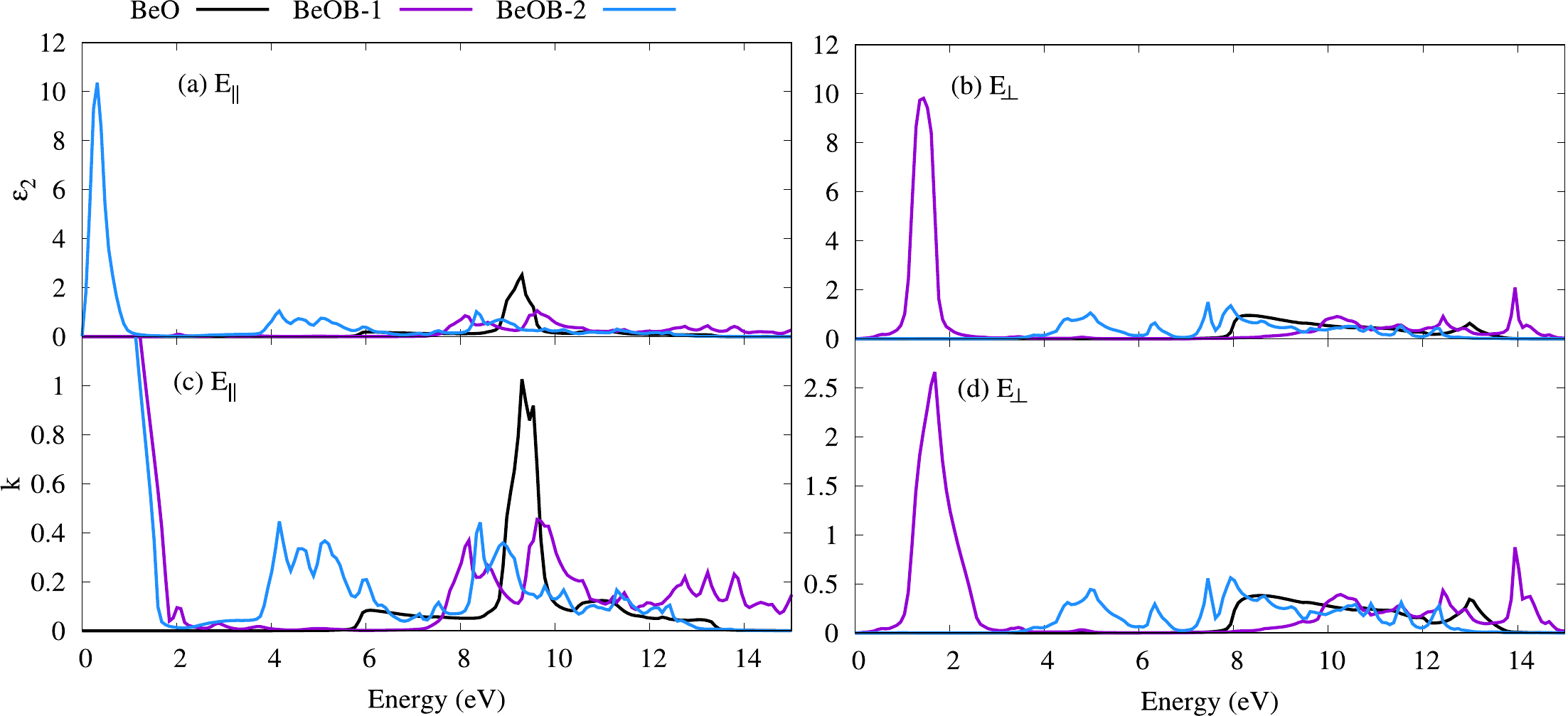}
	\caption{Imaginary part of dielectric function, $\varepsilon_2$ (a) and (b), and excitation spectra, $k$ (c) and (d), for the  $E_{\rm \parallel}$ (left panel) and $E_{\rm \perp}$ (right panel) of the electric field, respectively.}
	\label{fig10}
\end{figure*}
From the $\varepsilon_2$ (a), and $k$ (b) spectra of BeOB-1 and BeOB-2 shown in \fig{fig10}, one can see that a strong peak at low energy is seen in addition to several small peaks in the high energy range. The strong peak at energy $0.32$~eV for BeOB-2 in $E_{\rm \parallel}$ indicates an interband transition along the M-K direction, which is due to the degenerate semiconductor behavior of the BeOB-2 monolayer. A peak at $1.44$~eV for BeOB-1 in $E_{\rm \perp}$ is found with almost the same intensity for BeOB-2 in $E_{\rm \parallel}$, but broader.

This intense peak can be referred to the interband transition along the K-$\Gamma$ or the $\Gamma$-M directions. The more pronounced peak at $9.3$~eV of pure BeO is suppressed in the case of BeOB-1 and BeOB-2 monolayers, which is due to the descreased band gap at the K points of both monolayer structures. We note that the peak intensity for BeOB-1 and BeOB-2 at the low energy range, $0\text{-}2$ meV, is much stronger, than that for the BeON-1 and BeON-2 monolayers in the same energy range (see and compare \fig{fig04} and \fig{fig10}). This implies that the BeOB-1 and BeOB-2 monolayers have a strong ability to absorb visible light.
Another difference in the $\varepsilon_2$ and the $k$ spectrum between the BeON and BeOB monolayers is the peaks in the high energy range from $11.5$ to $14$~eV. The peaks in the high energy range are much stronger for BeON compared to the BeOB monolayers indicating that the BeON in the high energy range has a strong ability to absorb light.
\begin{figure}[htb]
	\centering
	\includegraphics[width=0.45\textwidth]{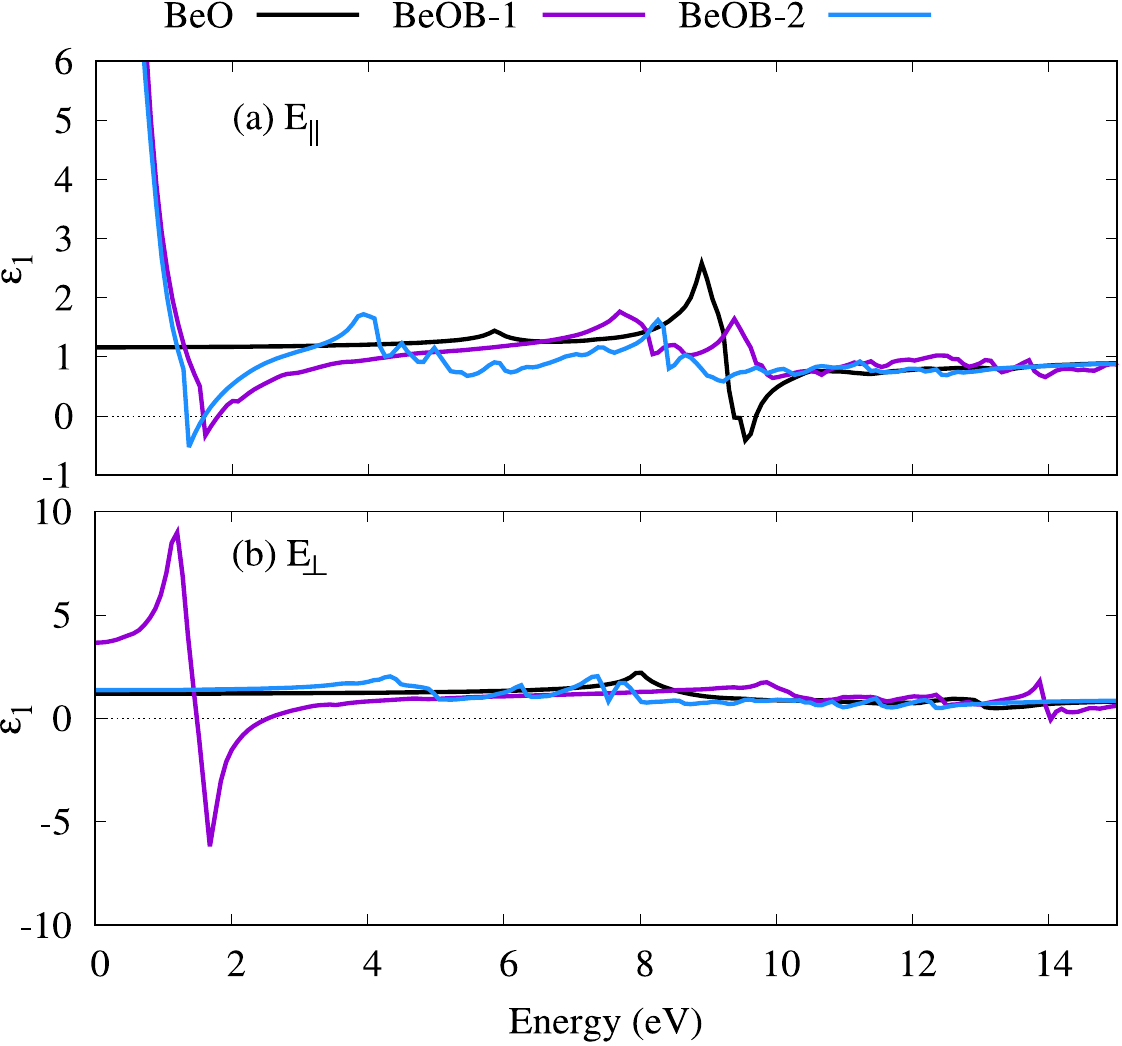}
	\caption{Real part of dielectric function, $\varepsilon_1$, for BeO (black), BeOB-1 (purple), and BeOB-2 (blue) in the case of $E_{\rm \perp}$.}
	\label{fig12}
\end{figure}

From the real dielectric function spectra of BeOB-1 and BeOB-2 monolayers shown in \fig{fig12}, one can see that the value of $\varepsilon_1(0)$ is also enhanced for BeOB-1 and BeOB-2 in the $E_{\rm \perp}$ (b). The $\varepsilon_1(0)$ is increased to $3.65$ and $1.36$ for BeOB-1 and BeOB-2, respectively. Furthermore, in several places of energy a zero value of $\varepsilon_1$ is visible. 
For instance, in the case of $E_{\rm \parallel}$, the value of $\varepsilon_1$ is zero at $1.32$ and $1.56$~eV for BeOB-2 and  BeOB-1, respectively, while it is zero at $1.49$~eV for BeOB-1 in the case of $E_{\rm \perp}$. We realize that the zero value of $\varepsilon_1$ at $1.49$~eV in the case of $E_{\rm \perp}$ gives rise to a plasmon frequency as the peak in $\varepsilon_2$ is formed at the same energy value.
\begin{figure}[htb]
	\centering
	\includegraphics[width=0.45\textwidth]{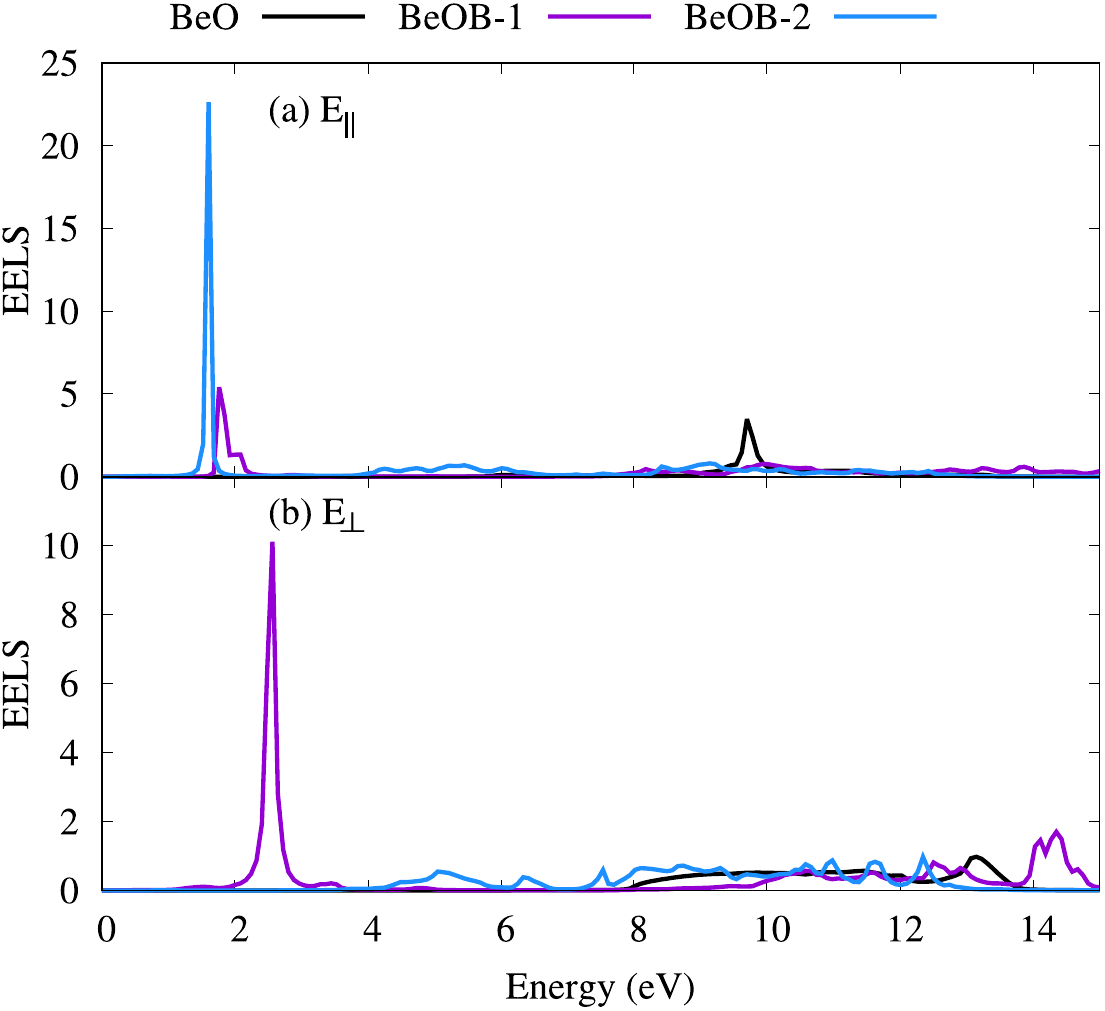}
	\caption{Electron energy loss spectrum, EELS, for BeO (black), BeOB-1 (purple), and BeOB-2 (blue) in the case of $E_{\rm \parallel}$ (a) and $E_{\rm \perp}$ (b) electric fields.}
	\label{fig11}
\end{figure}

The EELS of both the BeOB-1 and BeOB-2 monolayers in the case of $E_{\rm \parallel}$ (a) and $E_{\rm \perp}$ (b) electric field is displayed in \fig{fig11}. The first observation is that the EELS peak intensity of both BeOB monolayers is much stronger than that for the BeON monolayers at low energy range, $0\text{-}3$~eV. The intense peaks in this energy range of BeOB monolayers can be referred to the plasmon peaks caused by the valence/conduction electron density of a BeOB monolayer. This is expected as the B dopant atom has a higher electronegativity comparing to the N atom. As a result, valence/conduction electron density is increased leading to an intense peak in the low energy range. The second observation is that opposite to the low energy range the EELS peak of BeOB monolayers are much weaker than those of BeON in the high energy range, $12\text{-}14$~eV.

\begin{figure}[htb]
	\centering
	\includegraphics[width=0.45\textwidth]{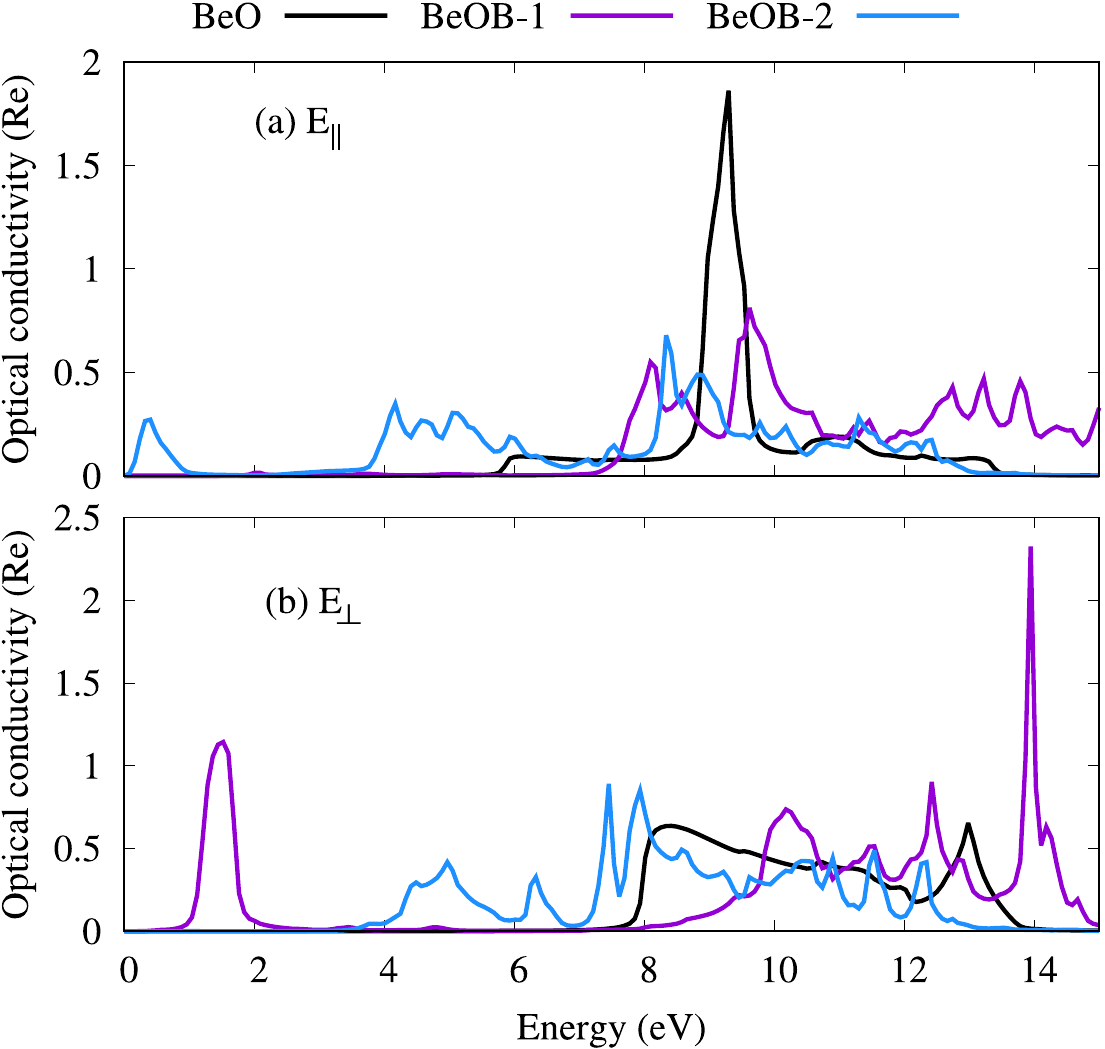}
	\caption{Optical conductivity (Real part), for BeO (black), BeOB-1 (purple), and BeOB-2 (blue) in the case of $E_{\rm \parallel}$ (a) and $E_{\rm \perp}$ (b) electric field.}
	\label{fig13}
\end{figure}

Finally, the real part of the optical conductivity of BeOB-1 (purple), and BeOB-2 (blue) is plotted for 
$E_{\rm \parallel}$ (a) and $E_{\rm \perp}$ (b) in\fig{fig13}. In the case of $E_{\rm \parallel}$, the optical conductivity for both B-doped monolayer is suppressed around the main intense peak of pure BeO monolayer due to modification of the band gap at the K point. But several weak peaks in the optical conductivity over the selected range of energy are found for both BeOB monolayers.

In contrast to $E_{\rm \parallel}$, two intense peaks in the optical conductivity of a BeOB-1 monolayer at low and high energy ranges are observed in the case of $E_{\rm \perp}$. 
The peak at low energy can be referred to interband transition and plasmon oscillations.

In the above sections, we see that in the N- or B-doped BeO monolayers the optical behavior 
is strongly modified compared to a pure BeO monolayers. This indicates the appearance of
promising properties, which could be useful for optoelectronic devices.

\section{Conclusion}\label{Sec:Conclusion}

In summary, the electronic and optical behaviors of boron- or nitrogen-doped beryllium
oxide monolayer have been studied by first principles calculations.
The electronic results show that N and B dopant atom can give rise to a
tunable band gap. We find that the effects of the N- or B-atoms on the band structure of the BeO monolayers
are very similar to the influences of these two atoms in graphene monolayer. 
In both cases, new electronic states around the Fermi energy are generated leading to the emergence of a degenerate semiconductor structure. 

We find that the charge transfers from the Be atoms to the N and B dopant atoms is caused by the higher electronegativity of the N and B atoms than the Be atoms. The charge modification in the N- or B-doped BeO can control the plasmon oscillation and the dielectric properties of the structures.
We observe that the value of the static dielectric constant is increased in the N- or B-doped monolayer compared to the pure BeO monolayer.
In addition, the imaginary part of the dielectric function, the excitation spectra, and the optical conductivity are enhanced in the low energy range, i.e. the visible range. The enhancement depends on the direction of the polarization of the incoming electric field.

Our results could be useful for improving the optoelectronic devices where BeO monolayers have been 
previously used .

\section{Acknowledgment}
This work was financially supported by the University of Sulaimani and 
the Research center of Komar University of Science and Technology. 
The computations were performed on resources provided by the Division of Computational 
Nanoscience at the University of Sulaimani.  
 
%\section{References}

%\bibliographystyle{elsarticle-num} 
%\bibliography{Ref_2.bib}

\end{document}